\documentclass[aps,prd,showpacs,amssymb,eqsecnum,amsfonts,epsf,
preprintnumbers,nofootinbib,superscriptaddress]{revtex4}
\usepackage[dvips]{graphicx}
\usepackage{bm,latexsym,amsmath,amssymb,amsfonts}
\usepackage{graphicx}
\usepackage{natbib}
\newcommand*{\cR}{{\cal R}}
\newcommand*{\cF}{{\cal F}}
\newcommand*{\cA}{{\cal A}}
\newcommand*{\cS}{{\cal S}}
\newcommand*{\cO}{{\cal O}}
\newcommand*{\e}{{\rm e}}
\begin{document}

\title{
Gravity on an extended brane in six-dimensional warped flux compactifications
}

\author{Tsutomu Kobayashi}
\email[Email: ]{tsutomu"at"th.phys.titech.ac.jp}
\affiliation{Department of Physics, Tokyo Institute of Technology, Tokyo 152-8551, Japan}
\author{Masato~Minamitsuji}
\email[Email: ]{masato"at"theorie.physik.uni-muenchen.de}
\affiliation{Arnold-Sommerfeld-Center for Theoretical Physics, Department f\"{u}r Physik, Ludwig-Maximilians-Universit\"{a}t, Theresienstr. 37, D-80333, Munich, Germany}

\begin{abstract}
We study linearized gravity in a six-dimensional Einstein-Maxwell model of warped braneworlds,
where the extra dimensions are compactified by a magnetic flux.
It is difficult to construct a strict codimension two braneworld
with matter sources other than pure tension.
To overcome this problem
we replace the codimension two defect by an extended brane,
with one spatial dimension compactified on a Kaluza-Klein circle.
Our background is composed of a warped, axisymmetric bulk and one or two branes.
We find that weak gravity sourced by arbitrary matter on the brane(s)
is described by a four-dimensional scalar-tensor theory.
We show, however, that the
scalar mode is suppressed at long distances and hence four-dimensional
Einstein gravity is reproduced on the brane. 
\end{abstract}

\preprint{LMU-ASC 15/07}
\maketitle

\section{Introduction}

Higher dimensional theories have been attracting much attention for many
years.
In the traditional Kaluza-Klein theories the compactification scale
must be microscopic to guarantee effectively four-dimensional (4D) spacetime.
In contrast to the Kaluza-Klein picture,
recently proposed braneworld scenarios~\cite{Rubakov}
allow for the presence of large (or even infinite) extra dimensions
owing to localization of matter fields on the brane.
Among various braneworld models, those with two extra dimensions
have received particular interests from a phenomenological point of view.
A naive reduction of six-dimensional (6D) gravity leads to
$M^2_{{\rm Pl}}=M^4{\cal V}$ \cite{Arkani-Hamed:1998rs},
where $M_{{\rm Pl}}$ is the observed 4D Planck mass and
$M$ is the fundamental scale of gravity, and ${\cal V}$ is the volume of extra dimensions.
For $M\sim 1~\text{TeV}$ we roughly have ${\cal V}^{1/2}\sim 1~\text{mm}$,
which eliminates the hierarchy problem in the usual sense
and at the same time offers us
the possibility to detect extra dimensions by table-top experiments on gravity.

In the present paper we focus on the 6D brane model
in which the internal two-dimensional space is compactified by a magnetic flux.
(See, e.g.,
Ref.~\cite{Papantonopoulos:2006uj} and references therein
for recent progresses in 6D braneworld models.)
The basic motivation for considering such models is that
extra dimensions, branes, and fluxes are important ingredients in string theory.
Braneworld settings with flux compactification
provide us natural toy models unifying aspects of string theory and cosmology,
though such scenarios have not been explored much.
Moreover, it has been suggested that the football (or rugby-ball) shaped codimension two braneworlds 
may help to resolve the cosmological constant problem 
by a self-tuning mechanism~\cite{cc}.
(The mechanism has been criticized for several reasons~\cite{cc2}.)

The behavior of gravity in 6D braneworlds should be investigated carefully.
In order for the brane scenario to be ``realistic'',
it is necessary to reproduce
4D Einstein gravity on the brane at least on scales much larger than
the compactification radius.
Properties of gravity in a codimension two braneworld with flux compactification
has been discussed so far in Ref.~\cite{Mukohyama}. However,
a codimension two brane suffers from the problem of the localization of matter.
Namely, it is difficult to put energy-momentum tensor
different from pure tension on the brane in
Einstein gravity~\cite{pt, Bostock}.
Actually it has been demonstrated that massless relativistic particles
can also be accommodated on a conical brane;
the true problem here lies in the fact that the gravitational radius
of a mass $m$ is enhanced on the brane and hence
gravity becomes nonlinear at distances much greater than a naive
estimate~\cite{shock}\footnote{
We wish to thank Nemanja Kaloper for pointing this out to us.}.
In any case the description of gravity on a conical defect is involved,
which significantly restricts the analysis of gravitational aspects of 6D braneworlds.
One way to evade this generic problem is
introducing codimension one brane
with one extra dimension compactified on a circle
(which we call an {\em extended brane})
instead of the original codimension two defect~\cite{Peloso1}.
(See Refs.~\cite{Bostock, Corradini, ish, Kaloper:2004cy, Kanno:2004nr,Charmousis:2005ey,deRham:2005ci}
for other ways out.)
Since the model has a Kaluza-Klein spatial direction in addition to one large
extra dimension, this construction is thought of
as a ``hybrid'' braneworld scenario~\cite{Carter:2006uk}.
(Very recently
this approach has been employed to regularize a conical singularity
in a different context~\cite{Kaloper}.)

Along this line, the authors of Ref.~\cite{Peloso1} were the first to study
the behavior of gravity in the presence of matter sources on the brane.
In their model the internal space is
football shaped (i.e., not warped) and
codimension two branes at the poles are
replaced by the extended branes wrapped around the axis of the football.
The perturbation analysis shows that
standard 4D gravity is reproduced on the extended brane~\cite{Peloso1}.
The system is stable under the most generic axisymmetric perturbations~\cite{Peloso2}.
In Ref.~\cite{PPZ} the same regularization technique
is used for
more general brane models with {\em warped} internal space
both in Einstein-Maxwell theory~\cite{Mukohyama} and in 6D
supergravity~\cite{Aghababaie:2003ar, 6d_sugra}.
In this paper we extend the work of~\cite{Peloso1} and 
investigate linearized gravity sourced by arbitrary matter
on extended branes in the Einstein-Maxwell model of
the warped braneworld~\cite{PPZ}.
As in~\cite{Peloso1}, we basically follow the well-developed analysis
of perturbations~\cite{GT} in the Randall-Sundrum braneworld~\cite{Randall:1999ee}.
This model will serve as a simple playground to study gravitational aspects of braneworld
scenarios in the context of string flux compactifications.

The paper is organized as follows. 
In Sec II we describe our background model.
Then we derive perturbation equations and boundary conditions in Sec.~III.
In Sec.~IV we discuss the behavior of linearized gravity on the brane.
Finally we draw our conclusions in Sec.~V.


\section{6D warped flux compactification and extended branes}

\subsection{Bulk solution with extended branes}

The 6D background model we will discuss was originally used to
compactify an internal two-dimensional space in Einstein-Maxwell theory~\cite{FR}.
We consider the system in the braneworld context, and
basically follow the construction of~\cite{Peloso1, PPZ}.
The bulk has a structure of Minkowski spacetime times a two-dimensional compact space, and
consists of $M+1$ regions (${\cal M}_I$) separated by $M$ extended branes ($\Sigma_i$).
Our action is
\begin{eqnarray}
S=\sum_IS_I+\sum_iS_i,
\end{eqnarray}
where the $I$-th bulk and $i$-th brane actions are given, respectively, by
\begin{eqnarray}
S_I&=&\int_{{\cal M}_I}
d^6x\sqrt{-g}\left[\frac{M^4}{2}\left(\cR-\frac{1}{L_I^2}\right)-\frac{1}{4}
  \cF_{MN}\cF^{MN}\right],
\\
S_i&=&-\int_{\Sigma_i} d^5x\sqrt{-q}
\left[\lambda_i+\frac{v_i^2}{2}q^{\hat\mu\hat\nu}
\left(\partial_{\hat \mu}\sigma_i -\e \cA_{\hat\mu}\right)
\left(\partial_{\hat \nu}\sigma_i -\e \cA_{\hat\nu}\right)\right].
 \label{branes}
\end{eqnarray}
Here the capital Latin indices $\{M,N,\cdots\}$ and the Greek indices
with hat $\{{\hat \mu},{\hat \nu},\cdots\}$ are used for
tensors defined in the 6D bulk and on the 5D brane(s), respectively.
We have the $U(1)$ gauge field $\cA_M$ with field strength $\cF_{MN}=\partial_M\cA_N-\partial_N\cA_M$
in the bulk, which is coupled to the brane scalar fields $\sigma_i$.
On each brane $\sigma_i$ arises as a Goldstone mode of a Higgs field, and the absolute value
of the vacuum expectation value of the Higgs field is given by $v_i/\sqrt{2}$.
We also include tension $\lambda_i$ in the brane actions\footnote{The
brane action of the form~(\ref{branes}) may not be unique
but rather is the simplest one.
See Ref.~\cite{Aghababaie:2003ar} for brane actions.}.
In this model the cosmological constant $1/(2L_I^2)$ is different at different regions in the bulk.

The background solution is described by the metric
\begin{eqnarray}
g_{MN}dx^Mdx^N =a^2(w)\eta_{\mu\nu}dx^{\mu}dx^{\nu}+L_{I}^2\left[\frac{dw^2}{f(w)}+\beta_I^2f(w)d\phi^2\right],
\end{eqnarray}
where
\begin{eqnarray}
f\left(w\right)=\frac{1}{5(1-\alpha)^2}\left[-a^2(w)
+\frac{1-\alpha^8}{1-\alpha^3}\frac{1}{a^3(w)}
-\frac{\alpha^3(1-\alpha^5)}{1-\alpha^3}\frac{1}{a^6(w)}
\right],
\end{eqnarray}
and the field strength
\begin{eqnarray}
\cF_{w\phi}=\frac{\mu M^2\beta_IL_I}{a^4}, \;\;
\mu:=\sqrt{\frac{3\alpha^3(1-\alpha^5)}{5(1-\alpha^3)}}. 
\end{eqnarray}
The warp factor is given by
\begin{eqnarray}
a(w)=\frac{1}{2}\left[(1-\alpha)w+1+\alpha\right],
\end{eqnarray}
and $\alpha \;\;(0<\alpha\leq 1)$ characterizes warping of the bulk.
The function $f(w)$ has two positive roots $w=1$ ($a=1$) and $w=-1$ ($a=\alpha$), and hence we will consider
the space $-1\leq w\leq 1$ ($\alpha \leq a \leq 1$).
The original construction of the braneworld model of this type
is given by Mukohyama et al. in Ref.~\cite{Mukohyama},
in which codimension two defects are considered.
The stability analysis against classical perturbations
has been performed in their subsequent papers~\cite{Yoshiguchi, Sendouda}.
In Appendix~\ref{App:deri} we replicate
the detailed derivation of the bulk solution~\cite{Mukohyama, PPZ}.

We put brane(s) at $w=\bar w_i$.
Then the continuity of the $(\phi\phi)$ component of the induced metric implies
that $\beta_IL_I$ is continuous across the brane. Therefore we write
\begin{eqnarray}
\ell:= \beta_IL_I.
\label{beta-L-ell}
\end{eqnarray}
The points $w=\pm 1$ can be regarded as poles.
We assume that $\phi$ has period $2\pi$. In order to avoid conical singularities at the two poles, we impose
\begin{eqnarray}
\beta_+=\frac{20(1-\alpha)(1-\alpha^3)}
{5-8\alpha^3+3\alpha^8},
\qquad
\beta_-=\frac{20(1-\alpha^{-1})(1-\alpha^{-3})}
{5-8\alpha^{-3}+3\alpha^{-8}},
\end{eqnarray}
where $\beta_{\pm}$ denotes the value of $\beta_I$ in the region that contains the pole $w=\pm1$.
It is easy to see that $\beta_+\geq 1\geq \beta_->0$.
Now we check that the above bulk solution coincides with the model of~\cite{Peloso1}
in the unwarped limit $\alpha\to1$.
Indeed, we have $f\to1-w^2$, ${\cal F}_{w\phi}\to M^2\beta_IL_I$, and $\beta_{\pm}\to1$ as $\alpha\to1$.
With the coordinate transformation $w=\sin\theta$ we can reproduce
the bulk solution of~\cite{Peloso1}. In this unwarped limit the background has a $Z_2$ symmetry
across the equator ($w=0$) before introducing branes.

Let us look at the quantization condition for the flux.
Since $\cF_{w\phi}=\cA_{\phi}'$, where the prime stands for a derivative with respect to $w$,
we have
\begin{eqnarray}
&&\cA_{\phi}=\cA^{(N)}_{\phi}:=-\frac{2\mu\ell M^2}{3(1-\alpha)}
\left(\frac{1}{a^3}-1\right)
\quad \mbox{for}\;\;
w>w_e-\varepsilon,
\label{bsolAN}\\
&&\cA_{\phi}=\cA^{(S)}_{\phi}:=-\frac{2\mu\ell M^2}{3(1-\alpha)}
\left(\frac{1}{a^3}-\frac{1}{\alpha^3}\right)
\quad \mbox{for}\;\;
w<w_e+\varepsilon,
\label{bsolAS}
\end{eqnarray}
where $w_e$ is arbitrary and $\varepsilon>0$.
The integration constants here are chosen so that $\cA_{\phi}=0$ at the poles.
We see that in the overlapping region $w_e-\varepsilon<w<w_e+\varepsilon$,
$\cA_{\phi}$ is doubly defined.
This means that $\cA^{(N)}_{\phi}$ and $\cA^{(S)}_{\phi}$
are related via the following gauge transformation:
\begin{eqnarray}
\cA^{(S)}_{\phi}=\cA^{(N)}_{\phi}+\partial_{\phi}\left(\mu_*\ell M^2\;\phi\right),\qquad
\mu_*:=\sqrt{\frac{(1-\alpha^3)(1-\alpha^5)}{15\alpha^3(1-\alpha)^2}}.
\end{eqnarray}
Noting that $\sigma_i$ is the phase of the brane Higgs field with charge $\e$ under the $U(1)$ symmetry,
we arrive at the following quantization condition:
\begin{eqnarray}
2\e\; \mu_*\ell M^2=N,\quad N=0,1,2,\cdots.
\end{eqnarray}

\begin{figure}[t]
  \begin{center}
    \includegraphics[keepaspectratio=true,height=80mm]{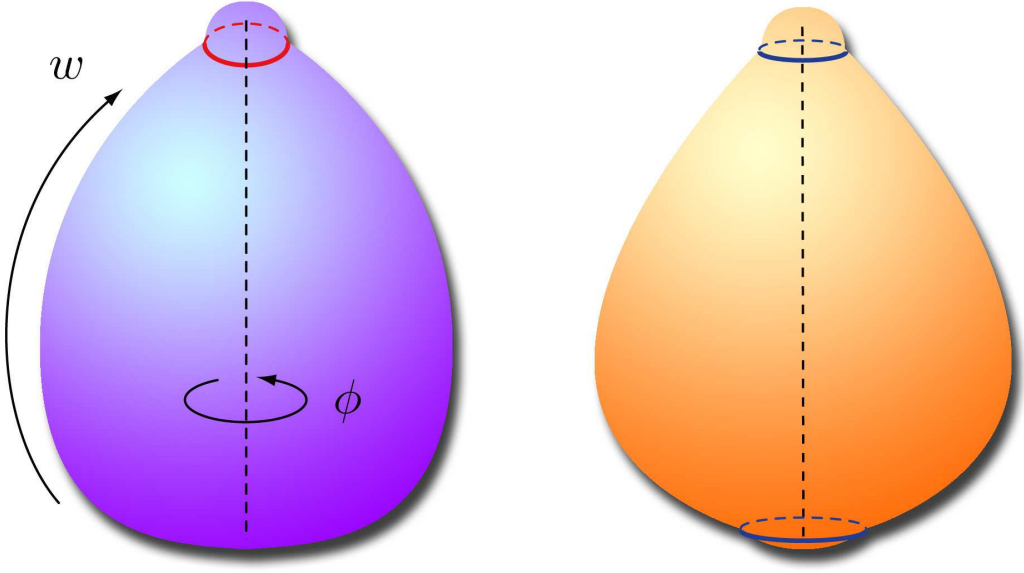}
  \end{center}
  \caption{Single brane and two-brane models.}%
  \label{fig:background.eps}
\end{figure}

The junction equations at a brane are given as follows.
They are derived from (i) the continuity of the induced metric, (ii) the Israel conditions, and
(iii) the junction conditions for the Maxwell field.
The first one was already used to relate $\beta_I$ with $L_I$ in Eq.~(\ref{beta-L-ell}).
The Israel conditions are written as
\begin{eqnarray}
\left[\left[
K_{\hat\mu\hat\nu}-q_{\hat\mu\hat\nu}K
\right]\right]_{\bar w_i}=-\frac{S^{(i)}_{\hat\mu\hat\nu}}{M^4},
\end{eqnarray}
where 
$S_{\hat\mu\hat\nu}$ is the energy-momentum tensor on the brane
and
$[[F]]_{\bar w_i}:=\lim_{\epsilon\to 0}\left(F|_{\bar w_i+\epsilon}-F|_{\bar w_i-\epsilon}\right)$.
It is straightforward to compute the extrinsic curvature.
The energy-momentum tensor on the brane is given by
\begin{eqnarray}
S_{\mu\nu}^{(i)}&=&-\left.\left[\lambda_i
+\frac{v^2_i}{2}q^{\phi\phi}\left(\partial_{\phi}\sigma_i-\e\cA_{\phi}\right)^2
\right]q_{\mu\nu}\right|_{\bar w_i},
\\
S_{\phi\phi}^{(i)}&=&-\left.\left[\lambda_i
-\frac{v^2_i}{2}q^{\phi\phi}\left(\partial_{\phi}\sigma_i-\e\cA_{\phi}\right)^2
\right]q_{\phi\phi}\right|_{\bar w_i},
\end{eqnarray}
Then the Israel conditions are rearranged to give
\begin{eqnarray}
~[[\beta_I]]_{\bar w_i}\left.\cdot\frac{f^{1/2}}{\ell}
\left(\frac{f'}{2f}- \frac{a'}{a} \right)\right|_{\bar w_i}&=&\left.-\frac{v_i^2}{M^4}
q^{\phi\phi}\left(\partial_{\phi}\sigma_i-\e\cA_{\phi}\right)^2\right|_{\bar w_i},
\label{Is1}
\\
~[[\beta_I]]_{\bar w_i}\left.\cdot\frac{f^{1/2}}{\ell}
\left(\frac{f'}{4f}+ \frac{7}{2}\frac{a'}{a} \right)\right|_{\bar w_i}
&=&-\frac{\lambda_i}{M^4}.
\end{eqnarray}

The junction conditions for the Maxwell field can be written as
\begin{eqnarray}
\left[\left[n^N\cF_{NM}\right]\right]_{\bar w_i}=
-\left.\e v_i^2\left(\partial_{M}\sigma_i-\e\cA_{M}\right)\right|_{\bar w_i},
\end{eqnarray}
where $n^N$ is the unit normal to the brane.
For the background, only the $\phi$ component is relevant:
\begin{eqnarray}
~[[\beta_I]]_{\bar w_i}\cdot \left.\frac{\mu M^2 f^{1/2}}{a^4}\right|_{\bar w_i}
=-\left.\e v_i^2\left(\partial_{\phi}\sigma_i-\e\cA_{\phi}\right)\right|_{\bar w_i}.
\label{maxwell_j_b}
\end{eqnarray}
From Eqs.~(\ref{Is1}) and~(\ref{maxwell_j_b}) we obtain
\begin{eqnarray}
[[\beta_I]]_{\bar w_i} = -\left.\ell\left(\frac{\e v_i}{\mu}\right)^2
\cdot a^8f^{1/2}\left(\frac{f'}{2f}-\frac{a'}{a}\right)\right|_{\bar w_i}.
\label{position_constraint}
\end{eqnarray}
Since $f'/(2f)-a'/a=0$ has a root $w_c$ ($-1<w_c\leq0$),
we see that
$[[\beta_I]]_{\bar w_i}\geq0$ for $\bar w_i\geq w_c$ and $[[\beta_I]]_{\bar w_i}<0$ for $\bar w_i < w_c$.

The equation of motion for $\sigma_i$ implies
$\partial^{\phi}\left(\partial_{\phi}\sigma_i-\e\cA_{\phi}\right)=0$,
leading to
$\sigma_i=n_i \phi$,
where $n_i$ is an integer.
Using Eqs.~(\ref{Is1}) and~(\ref{maxwell_j_b}) we obtain~\cite{PPZ}
\begin{eqnarray}
n_i=-\frac{N}{1-\alpha^3}\left[
\frac{5(1-\alpha^8)}{8(1-\alpha^5)}-\alpha^3c_i
\right],
\end{eqnarray}
where $c_i=1$ for $\bar w_i>w_e$ and $c_i=\alpha^{-3}$ for $\bar w_i<w_e$
[see Eqs.~(\ref{bsolAN}) and~(\ref{bsolAS})].
In the limit $\alpha\to1$, we can see that $n_i=-N/2$
for $\bar w_i>w_e$ and $n_i=+N/2$ for $\bar w_i<w_e$~\cite{Peloso1}.

\subsection{Single brane model}

\begin{figure}[tb]
  \begin{center}
    \includegraphics[keepaspectratio=true,height=60mm]{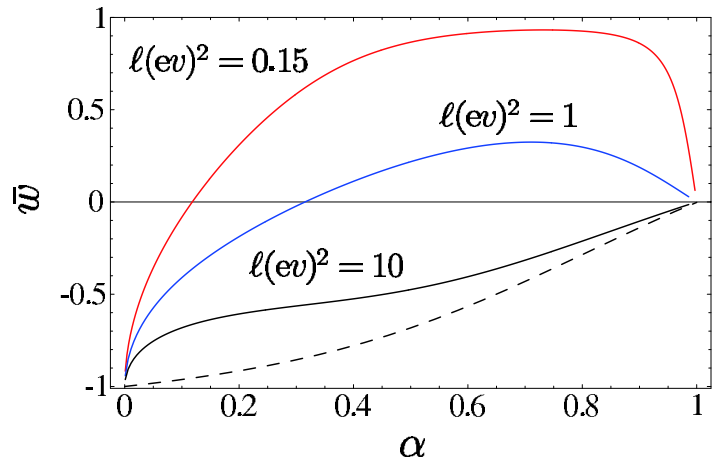}
  \end{center}
  \caption{The position of the brane in the single brane model. The dashed line shows the value $w_c$
  and hence one cannot put a brane below this line.}
  \label{fig:1brane_position.eps}
\end{figure}

Let us consider a model with a single brane at $w=\bar w$.
The two regions separated by the brane have the curvature scales
$L_+=\ell/\beta_+$ and $L_-=\ell/\beta_-$, respectively.
The configuration of the model is described in the left side of Fig.~\ref{fig:background.eps}.
For given parameters $\alpha$ and $\ell(\e v)^2$,
Eq.~(\ref{position_constraint}) determines the position of the brane $\bar w$, as
is shown in Fig.~\ref{fig:1brane_position.eps}.
Note that one cannot put the brane in the region $w<w_c$ because of the condition $\beta_+\geq\beta_-$.

\subsection{Two-brane model}

We can also consider a two-brane model, which is
nothing but the situation considered in Ref.~\cite{PPZ}.
In this case, the bulk is divided into three parts by two branes at $w=\bar w_+$ and $w=\bar w_-$.
A schematic picture is shown in the right side of Fig.~\ref{fig:background.eps}.
The three bulk regions are characterized by the curvature scales
$L_+=\ell/\beta_+$, $L_0:=\ell/\beta_0$, and $L_-=\ell/\beta_-$,
respectively, from north to south.
If $w_c>\bar w_+>\bar w_-$, it follows from
Eq.~(\ref{position_constraint}) that $\beta_+<\beta_0<\beta_-$.
This is incompatible with the fact
$\beta_+\geq\beta_-$, and so we impose $\bar w_+\geq w_c$.
Consequently, we have $\beta_+\geq \beta_0$.
However, both $\bar w_-\geq w_c$ ($\beta_0\geq\beta_-$) and
$\bar w_-<w_c$ ($\beta_0<\beta_-$) are possible.

The background configuration of~\cite{Peloso1} can be recovered by taking the unwarped limit ($\alpha\to1$)
and assuming at the same time that $v_+=v_-$.
As is mentioned in the above, the unwarped bulk automatically
has the $Z_2$-symmetry across the equator at $w=0$.
The second condition requires that the branes are located at $Z_2$-symmetric positions, i.e., $\bar w_-=-\bar w_+$.
Furthermore,
in Ref.~\cite{Peloso1} the $Z_2$-symmetry is imposed {\em also for linear perturbations}.
We stress that the general warped background ($\alpha\neq1$) has no such symmetry.
Our perturbation analysis in the rest of the paper does include
the unwarped model as the limiting case, but we will not restrict ourselves
to the $Z_2$-symmetric brane configuration nor
perturbations in the limit $\alpha\to1$.
We will comment on this point in more detail in Sec.~\ref{phimode}.


\section{Linear perturbations}

We now consider linear perturbations from the background solution described in the
previous section.
We write the perturbed metric in an arbitrary gauge as
\begin{eqnarray}
\left(g_{MN}+\delta g_{MN}\right)dx^Mdx^N
&=&a^2\left[(1+2\Psi)\eta_{\mu\nu}+2E_{,\mu\nu}
+E_{\mu,\nu}+E_{\nu,\mu}+h_{\mu\nu}\right]
dx^{\mu}dx^{\nu}
\nonumber\\&&\quad
+2\left(B_{w,\mu}+B_{w\mu}\right)dwdx^{\mu}
+2\left(B_{,\mu}+B_{\mu}\right)d\phi dx^{\mu}
\nonumber\\&&\qquad
+L^2_I\left[
(1+2W)\frac{dw^2}{f}+2C \beta^2_Ifdwd\phi + (1+2\Phi)\beta^2_Ifd\phi^2
\right],
\end{eqnarray}
where the perturbations are split into scalar, vector, and tensor modes
under the Lorentz group in the external spacetime.
The vector modes $E_{\mu}$, $B_{w\mu}$, and $B_{\mu}$ are transverse:
$E_{\mu}^{~,\mu}=0, \cdots$. The tensor mode $h_{\mu\nu}$
is transverse and traceless: $h_{\mu\nu}^{\quad,\nu}=0=h_{\mu}^{~\mu}$.
Similarly, we can write the perturbations of the gauge field $\delta\cA_M$ as
\begin{eqnarray}
\delta\cA_{\mu}=\partial_{\mu}A+\hat A_{\mu},\quad
\delta\cA_{w}=A_w,\quad
\delta\cA_{\phi}=A_{\phi},
\end{eqnarray}
where $\hat A_{\mu}$ is a vector mode and the other three are scalar modes.
The scalar part of the perturbed field strength $\delta\cF_{MN}$ is given by
\begin{eqnarray}
\delta\cF_{w\phi}=A_{\phi}',
\quad
\delta\cF_{\mu\phi}=A_{\phi,\mu},
\quad
\delta\cF_{\mu w}=F_{w,\mu},
\end{eqnarray}
where
$F_w:=A_w-A'$.
The vector part of the perturbed field strength is
\begin{eqnarray}
\hat{\delta\cF}_{w\mu}=\hat A_{\mu}',
\quad
\hat{\delta \cF}_{\mu\nu}=\partial_{\mu}\hat A_{\nu}-\partial_{\nu}\hat A_{\mu}.
\end{eqnarray}
We shall discuss tensor and scalar perturbations in the following subsections.
We show in Appendix~\ref{App:vcmode} that
there are no vector-type perturbations of importance.

\subsection{Tensor mode}

Tensor perturbations are invariant under an infinitesimal coordinate transformation.
The 6D equation of motion for tensor perturbations is
\begin{eqnarray}
\left(a^4fh_{\mu\nu}'\right)'+a^2L^2_I\Box h_{\mu\nu}=0,\label{EOM-T}
\end{eqnarray}
where $\Box:=\eta^{\mu\nu}\partial_{\mu}\partial_{\nu}$.
Setting $\Box=0$ we obtain the zero-mode solution
\begin{eqnarray}
h_{\mu\nu}(x, w)={\mathtt A}^{I}_{\mu\nu}(x)+{\mathtt B}^{I}_{\mu\nu}(x)\int^w \frac{dw}{a^4 f},
\end{eqnarray}
where ${\mathtt A}^I_{\mu\nu}$ and ${\mathtt B}^I_{\mu\nu}$ are integration constants.
Requiring the regularity of $h_{\mu\nu}$ and $h_{\mu\nu}'$ at the poles~\cite{Yoshiguchi, Sendouda},
we see that ${\mathtt B}_{\mu\nu}^I$ vanishes in the region that contains the pole.
Thus, in the single brane model the zero mode is given by $h_{\mu\nu}={\mathtt A}^I_{\mu\nu}(x)$ everywhere.
The source-free Israel condition reads
$[[\beta_Ih_{\mu\nu}']]_{\bar w_i}=0$.
This means that, also in the two-brane model, the zero mode is written as
$h_{\mu\nu}={\mathtt A}^I_{\mu\nu}(x)$ everywhere.
In the next section,
we will further discuss the tensor mode with the introduction of matter sources on the brane(s).

\subsection{Scalar modes}


This section focuses on the master equations and boundary conditions
governing the dynamics of scalar-type variables.
The details of the computations of the Einstein and Maxwell equations
are given in Appendix~\ref{gts}.

As shown in the case of the codimension two model~\cite{Yoshiguchi, Sendouda},
the perturbation equations in the longitudinal gauge\footnote{
Scalar-type variables with tilde denote perturbations
in the longitudinal gauge, which is explained in Appendix~\ref{gts}.},
~(\ref{ww})--(\ref{tracemunu}) and (\ref{maxwell_phi}),
can be solved using two master variables defined by
\begin{eqnarray}
\Omega_1:=-\tilde \Phi-3\tilde \Psi,\;\;\Omega_2:=\tilde\Psi.
\end{eqnarray}
With a straightforward computation
Eqs.~(\ref{ww})--(\ref{tracemunu}) and (\ref{maxwell_phi})
can be recast into a set of coupled equations
\begin{eqnarray}
\Omega_1''+2\left(\frac{f'}{f}+5\frac{a'}{a}\right)\Omega_1'-\frac{2}{f}\left(\Omega_1+\Omega_2\right)
+\frac{L_I^2}{a^2f}\Box\Omega_1&=&0,\label{mas1}
\\
\Omega_2''+4\frac{a'}{a}\Omega_2'+\frac{L_I^2}{2a^2f}\Box\left(\Omega_1+2\Omega_2\right)&=&0.\label{mas2}
\end{eqnarray}
In terms of these master variables, the remaining metric and gauge field perturbations
are given by
\begin{eqnarray}
\tilde W&=&\Omega_1+\Omega_2,\\
\frac{\mu}{\ell M^2a^4}\tilde A_{\phi}&=&
f\left[\Omega_1'+\frac{f'}{f}\left(\Omega_1+2\Omega_2\right)+2\frac{a'}{a}\Omega_1\right].
\end{eqnarray}
We have three more variables, $\tilde C$, $\tilde A$, and $\tilde A_w$.
As explained in Appendix~\ref{App:vm} we do not have to take care of these modes
in any case.

We are interested in the zero-mode sector because of its importance
in describing the long range gravitational force.
Setting $\Box=0$ in Eqs.~(\ref{mas1}) and~(\ref{mas2}), we obtain analytic solutions
for the zero modes:
\begin{eqnarray}
\Omega_1&=&\frac{1}{5(1-\alpha)^2f}\left[
\frac{p_I(x)}{a^3}+\frac{q_I(x)}{a^6}
+u_I(x)a^2+\frac{4v_I(x)}{a}\right],
\label{zsol_1}\\
\Omega_2&=&u_I(x)+\frac{v_I(x)}{a^3},
\label{zsol_2}
\end{eqnarray}
where the integration constants $p_I, q_I, u_I, v_I$ are to be determined by the boundary conditions. 
In the unwarped $\alpha=1$ case, general solutions for the zero modes
are\footnote{Eqs.~(\ref{zsol_1}) and~(\ref{zsol_2}) do not coincide with
Eqs.~(\ref{zsola1}) and~(\ref{zsola2}) in the limit of $\alpha\to1$. However,
by replacing the coefficients $u_I$ and $v_I$ as
$u_I\to u_I+[(5-3\alpha)/3(1-\alpha)]v_I$ and $v_I\to-[2/3(1-\alpha)]v_I$
we obtain an expression for $\Omega_2$ that
reduces to Eq.~(\ref{zsola2}) in the unwarped limit.
We can obtain such an expression for $\Omega_1$
by replacing $p_I$ and $q_I$ in Eq.~(\ref{zsol_1}) with appropriate combinations
of $p_I, q_I, u_I$, and $v_I$.
The resultant expression is unnecessarily complicated.
}
\begin{eqnarray}
\Omega_1&=&\frac{1}{1-w^2}\left[p_I(x)+q_I(x) w+u_I(x)w^2+\frac{1}{3}v_I(x)w^3\right]\label{zsola1},
\\
\Omega_2&=&u_I(x)+v_I(x)w.\label{zsola2}
\end{eqnarray}

Since $f=0$ at the poles and $\Omega_1$ should be regular there,
we impose
\begin{eqnarray}
\left.f\Omega_1\right|_{w=\pm 1}=0.\label{bcp1}
\end{eqnarray}
We also require that $\tilde A_{\phi}=0$ at the poles, i.e.,
\begin{eqnarray}
\left[\left(f\Omega_1\right)'+2f'\Omega_2\right]_{w=\pm1} = 0.\label{bcp2}
\end{eqnarray}
A more rigorous argument on the boundary conditions at the poles is found in Ref.~\cite{Sendouda}.

We now turn to the boundary conditions at the branes.
The continuity of the induced metric and the gauge field imposes
\begin{eqnarray}
\left[\left[\tilde\Psi+\frac{a'}{a}\zeta\right]\right]_{\bar w_i}&=&0,\label{con1}
\\
\left[\left[\tilde\Phi+\frac{f'}{2f}\zeta\right]\right]_{\bar w_i}&=&0,\label{con2}
\\
\left[\left[\tilde A_{\phi}+\cA_{\phi}'\zeta\right]\right]_{\bar w_i}&=&0,\label{con3}
\end{eqnarray}
where we simply write the bending mode of each brane as $\zeta^{(i)}=\zeta$.

The $(\mu\nu)$ component of the Israel conditions, including the transverse and traceless perturbation, is
\begin{eqnarray}
\left[\left[
\frac{L_I}{f^{1/2}}\left(\zeta_{,\mu\nu}-\Box\zeta\eta_{\mu\nu}\right)
-\frac{f^{1/2}}{2L_I }a^2 h_{\mu\nu}'
\right]\right]_{\bar w_i}
= \frac{T_{\mu\nu}^{(i)}}{M^4},
\label{Is_munu}
\end{eqnarray}
where $T_{\mu\nu}^{(i)}$ is the matter energy-momentum tensor on the brane labeled by $i$. 
The four-dimensional trace of Eq.~(\ref{Is_munu}) reduces to
\begin{eqnarray}
\left[\left[
\frac{L_I }{f^{1/2}}\frac{1}{a^2}\Box\zeta
\right]\right]_{\bar w_i}
= -\frac{1}{3}\frac{T_{\lambda}^{~\lambda(i)}}{M^4}.
\label{Is_trace}
\end{eqnarray}
The $(\phi\phi)$ component of the Israel conditions gives
\begin{eqnarray}
\left[\left[
\frac{f^{1/2}}{L_I}\left\{-4\tilde\Psi'+4\frac{a'}{a}\tilde W
+\left(\frac{f'}{2f}-\frac{a'}{a}\right)\tilde\Phi+\frac{L_I^2}{a^2f}\Box\zeta+
\tilde Y\right\}
\right]\right]_{\bar w_i}
=-\frac{T_{\phi}^{~\phi(i)}}{M^4},
\label{Is_phiphi}
\end{eqnarray}
where we defined
\begin{eqnarray}
\tilde Y:=\left(\frac{f'}{2f}-\frac{a'}{a}\right)\left(\frac{f'}{2f}-4\frac{a'}{a}\right)\zeta
+\frac{\mu^2}{a^8f}\zeta+\frac{\mu}{\ell M^2a^4f}\tilde A_{\phi}.
\end{eqnarray}
The $\phi$ component of the Maxwell junction conditions gives
\begin{eqnarray}
\left[\left[
\beta_I\left\{
\left(\frac{f'}{2f}-\frac{a'}{a}\right)\left(\tilde A_{\phi}'-\frac{\mu\ell M^2}{a^4}\tilde W\right)
+\frac{\mu\ell M^2}{a^4}\tilde Y\right\}
\right]\right]_{\bar w_i}=0.\label{phimax}
\end{eqnarray}


\section{Linearized gravity on the brane}\label{lingr}

\subsection{The zero-mode truncation}

Following the approach of~\cite{GT} we shall see the behavior of weak gravity
created by matter sources on the brane(s).
Rearranging the Israel condition~(\ref{Is_munu}) we obtain
\begin{eqnarray}
\left[\left[\beta_Ia^4fh_{\mu\nu}'\right]\right]_{\bar w_i}=-\cS_{\mu\nu}^{(i)},\label{bct}
\end{eqnarray}
where we collected together the matter sources and the brane bending scalars
in the right hand side:
\begin{eqnarray}
\cS_{\mu\nu}^{(i)}:=
2a^2_i\left\{\frac{\ell f^{1/2}_i }{M^4}\left(T^{(i)}_{\mu\nu}
-\frac{T_{\lambda}^{~\lambda(i)}}{3}q_{\mu\nu}^{(i)}\right)
-\ell^2\left[\left[\beta^{-1}_I\zeta_{,\mu\nu}\right]\right]_{\bar w_i}\right\},
\end{eqnarray}
with $a_i:=a(\bar w_i)$ and $f_i:=f(\bar w_i)$.
We can now put the equation of motion~(\ref{EOM-T}) and the boundary condition~(\ref{bct})
into a single equation with source terms as
\begin{eqnarray}
\cO h_{\mu\nu}=-\sum_i\cS_{\mu\nu}^{(i)}\delta(w-\bar w_i),\label{EOM2}
\end{eqnarray}
where we defined an operator
\begin{eqnarray}
\cO h_{\mu\nu} :=\beta_I\left[a^4 f h_{\mu\nu}'\right]'+a^2\ell^2\beta^{-1}_I\Box h_{\mu\nu}.
\end{eqnarray}
Then using the (retarded) Green function defined by
\begin{eqnarray}
\cO G_R(x,w;x',w') = \delta^{(4)}(x-x')\delta(w-w'),
\end{eqnarray}
we can solve Eq.~(\ref{EOM2}):
\begin{eqnarray}
h_{\mu\nu}(x,w)=-\sum_i\int d^4x'\;G_R(x,w;x',\bar w_i)\cS_{\mu\nu}^{(i)}.
\end{eqnarray}
The Green function is explicitly given by
\begin{eqnarray}
G_R(x,w;x',w')=-\int \frac{d^4k}{(2\pi)^4}e^{ik\cdot(x-x')}\sum_n
\frac{\psi_n(w)\psi_n(w')}{m_n^2+\mathbf{k}^2-(\omega+i\epsilon)^2},
\end{eqnarray}
where $\psi_n(w)$ are a complete set of eigenfunctions of
\begin{eqnarray}
[a^4 f \psi_n']'=-a^2L^2_I m_n^2 \psi_n
\end{eqnarray}
and are normalized as
\begin{eqnarray}
\ell^2\int_{-1}^1\psi_n(w)\psi_{n'}(w)a^2(w) \beta_I^{-1} dw = \delta_{nn'}.
\end{eqnarray}

Here we are interested, for example, in gravity acting on isolated sources separated by
a distance much larger than the bulk length scale $\ell$.
In such situations, we may expect that the effect of Kaluza-Klein modes is suppressed
so that we are justified to truncate the Green function by keeping only the zero-mode contribution.
As already shown in the previous section
the zero-mode solution is constant,
$\psi_0=\ell_*^{-2}$, where
\begin{eqnarray}
\ell_*:=\ell\left[\int^1_{-1}\beta_I^{-1}a^2(w)dw\right]^{1/2}.
\end{eqnarray}
Note that we basically have $\ell_*\sim O(\ell)$.
The zero-mode truncation of the Green function yields
\begin{eqnarray}
h_{\mu\nu}\approx h_{\mu\nu}^{(m)}+h_{,\mu\nu}^{(\zeta)},
\end{eqnarray}
where we have separated $h_{\mu\nu}$ into the ``matter'' and ``brane bending'' contributions as
\begin{eqnarray}
h_{\mu\nu}^{(m)}&:=&-\frac{2\ell}{\ell_*^2 M^4}\sum_{i}a^2_if^{1/2}_i
\Box^{-1}\left(T_{\mu\nu}^{(i)}
-\frac{T_{\lambda}^{~\lambda(i)}}{3}q^{(i)}_{\mu\nu}\right),
\\
h^{(\zeta)}&:=&\frac{2\ell^2}{\ell_*^2}\sum_{i}a^2_i \Box^{-1}\left[\left[
\beta_I^{-1}\zeta\right]\right]_{\bar w_i}.
\end{eqnarray}

In order to discuss the behavior of gravity on the brane, we compute the linearized 4D Ricci tensor
for the induced metric
\begin{eqnarray}
\bar{q}_{\mu\nu}^{(i)}=\left.a^2_i\left[\left(1+2\bar\Psi\right)\eta_{\mu\nu}
+h_{\mu\nu}\right]\right|_{\bar w_i},
\end{eqnarray}
where $\bar\Psi$ is a metric perturbation in the Gaussian-normal gauge.
We can write the Ricci tensor as
\begin{eqnarray}
\bar R_{\mu\nu}^{(i)}
&=&-\frac{1}{2}\Box h_{\mu\nu}^{(m)}+\frac{1}{4}\Box^2h^{(\zeta)}\eta_{\mu\nu}
-\left(\partial_{\mu}\partial_{\nu}+\frac{1}{2}\eta_{\mu\nu}\Box\right)\varphi^{(i)},
\label{rtonb}
\end{eqnarray}
where
\begin{eqnarray}
\varphi^{(i)}&:=&
\left. \left(2\bar\Psi+\frac{1}{2}\Box h^{(\zeta)}\right)\right|_{\bar w_i}
=2\bar\Psi|_{\bar w_i}+\frac{\ell^2}{\ell^2_*}\sum_j a^2_j
\left[\left[\beta_I^{-1}\zeta\right]\right]_{\bar w_j}.
\label{bdsc}
\end{eqnarray}

Using the trace of the Israel condition~(\ref{Is_trace}), we find that
the first two terms give
\begin{eqnarray}
-\frac{1}{2}\Box h_{\mu\nu}^{(m)}+\frac{1}{4}\Box^2h^{(\zeta)}\eta_{\mu\nu}
&=&\sum_i 8\pi G^{(i)}\left[\left(\bar T_{\mu\nu}^{(i)}
-\frac{\bar T_{\lambda}^{~\lambda(i)}}{3}\bar{q}^{(i)}_{\mu\nu}\right)
-\frac{\bar T_{\lambda}^{~\lambda(i)}}{6}\bar{q}^{(i)}_{\mu\nu}\right]\nonumber
\\
&=&\sum_i 8\pi G^{(i)}\left(\bar T_{\mu\nu}^{(i)}
-\frac{\bar T_{\lambda}^{~\lambda(i)}}{2}\bar{q}^{(i)}_{\mu\nu}\right),
\end{eqnarray}
where $\bar T_{\mu\nu}$ is the energy-momentum tensor integrated along the $\phi$ direction,
\begin{eqnarray}
\bar T_{\mu\nu}^{(i)}:=\int_0^{2\pi}T^{(i)}_{\mu\nu}\sqrt{\bar q^{(i)}_{\phi\phi}}d\phi
=2\pi\ell f^{1/2}(\bar w_i)T_{\mu\nu}^{(i)},
\end{eqnarray}
and the 4D gravitational coupling at each brane is defined as
\begin{eqnarray}
8\pi G^{(i)}:=\frac{a^2(\bar w_i)}{2\pi \ell_*^2 M^4}.\label{4dpl}
\end{eqnarray}
Now it is clear how the 4D tensor structure can be recovered by the first two terms
in Eq.~(\ref{rtonb}); the brane bending plays a crucial role. The same mechanism works
in the Randall-Sundrum model~\cite{GT} and in the unwarped, $Z_2$-symmetric model of~\cite{Peloso1}.
What would make brane gravity different from 4D Einstein is the $\varphi^{(i)}$ term.
Apparently, gravity on the brane is described by a scalar-tensor theory due to this scalar mode.
However, as we shall show in the next subsection,
the effect of $\varphi^{(i)}$ term can be ignored as far as
long range gravity is concerned.

\subsection{The $\varphi^{(i)}$ mode}\label{phimode}

\begin{figure}[t]
  \begin{center}
    \includegraphics[keepaspectratio=true,height=70mm]{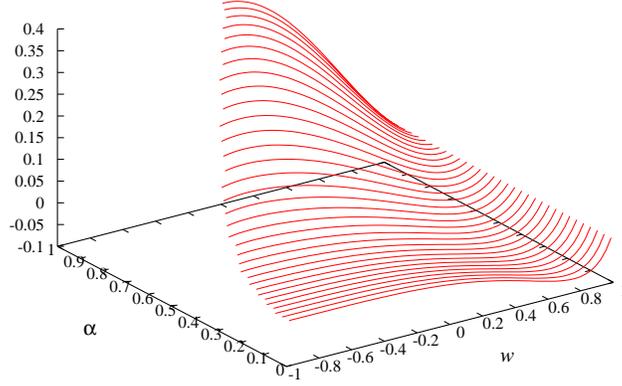}
  \end{center}
  \caption{Plot for $\Delta(\bar w, \alpha)$. The location of the brane
  is restricted in the region $\bar w\geq w_c$.}%
  \label{fig:plot_single.eps}
\end{figure}

We will now evaluate the impact of $\varphi^{(i)}$ on brane gravity and
make clear the physical meaning of this mode.
We start with combining Eqs.~(\ref{Is_trace}) and~(\ref{Is_phiphi}) to obtain
\begin{eqnarray}
\left[\left[
\beta_I\left\{-4\tilde\Psi'+4\frac{a'}{a}\tilde W
+\left(\frac{f'}{2f}-\frac{a'}{a}\right)\tilde\Phi+\tilde Y\right\}
\right]\right]_{\bar w_i}
=\frac{\ell^2_*}{a^2_i}8\pi G^{(i)}\bar\tau^{(i)},
\label{evaluate_tau}
\end{eqnarray}
where
\begin{eqnarray}
\bar \tau^{(i)} :=\frac{1}{3}\bar T_{\lambda}^{~\lambda(i)}-\bar T_{\phi}^{~\phi(i)},
\end{eqnarray}
and $\bar T_{\phi}^{~\phi(i)}$ is defined similarly to $\bar T_{\mu\nu}^{(i)}$.
This combination of the energy-momentum tensor is the key quantity.
To catch the meaning of $\varphi^{(i)}$, the following two quantities are important:
the perturbed volume of the internal space $\delta{\cal V}$
and the perturbed circumference of each brane $\delta{\cal C}^{(i)}$.
We can compute the volume of the internal space by performing the integral 
\begin{eqnarray}
{\cal V}_0+\delta {\cal V} = \int \sqrt{\bar g_{ww}\bar g_{\phi\phi}} dwd\phi,
\end{eqnarray}
yielding ${\cal V}_0 = 2\pi\ell^2\int\beta_I^{-1}dw$ and
\begin{eqnarray}
\delta{\cal V}=-2\pi\ell^2\left(
\sum_i\left[\left[\beta^{-1}_I\zeta\right]\right]_{\bar w_i}+2\int \beta^{-1}_I\Omega_2dw\right).
\end{eqnarray}
The circumference of the brane is given by
\begin{eqnarray}
{\cal C}^{(i)}_0+\delta{\cal C}^{(i)}=\int\sqrt{\bar g_{\phi\phi}}d\phi,
\end{eqnarray}
and so we have ${\cal C}_0^{(i)}= 2\pi\ell f^{1/2}_i$ and
\begin{eqnarray}
\delta{\cal C}^{(i)}&=&2\pi\ell f^{1/2}_i \left.\bar\Phi\right|_{\bar w_i}.
\end{eqnarray}

Let us first consider a single brane model. In this case
there are $4\times2=8$ scalar integration constants
and $2$ brane bending modes. Using $4$ boundary conditions at the brane
[Eqs.~(\ref{con1})--(\ref{con3}) and~(\ref{phimax})] and
$2\times2=4$ regularity conditions at the two poles [Eqs.~(\ref{bcp1}) and~(\ref{bcp2})],
we can express $8$ of $10$ variables in terms of $2$ variables, say $u_-(x)$ and $v_-(x)$.
Then, substituting the result into Eqs.~(\ref{bdsc}) and~(\ref{evaluate_tau})
we can express $\varphi$ and $\bar\tau$ in terms of $u_-(x)$ and $v_-(x)$.
This procedure reveals the relation
\begin{eqnarray}
\varphi = \frac{\ell_*^2}{a^2(\bar w)} 8\pi G\bar\tau\:\Delta(\bar w, \alpha),
\end{eqnarray}
where $\Delta(\bar w, \alpha)$ is a regular function and it is dimensionless.
Since the explicit form of $\Delta(\bar w, \alpha)$ is quite messy,
we show a plot in Fig.~\ref{fig:plot_single.eps} instead of writing down a lengthy equation.
Now we have
\begin{eqnarray}
\bar R_{\mu\nu}
= 8\pi G \left[\left(\bar T_{\mu\nu}
-\frac{\bar T_{\lambda}^{~\lambda}}{2}\bar{q}_{\mu\nu}\right)
-\Delta(\bar w, \alpha)\frac{\ell^2_*}{a^2(\bar w)}
\left(\partial_{\mu}\partial_{\nu}+\frac{1}{2}\eta_{\mu\nu}\Box\right)\bar \tau
\right].
\end{eqnarray}
From this it is easy to see that the second term is negligible on scales much larger than $\ell_*(\sim\ell)$.

The scalar mode $\varphi$ decouples from gravity on the brane
as the brane shrinks to the pole ($\bar w \to+1$),
since $\Delta\to 0$ in this limit.
This is in agreement with the result of~\cite{Peloso1}.
Note, however, that in fact one cannot bring the brane to the pole for a finite value of $v$
because of Eq.~(\ref{position_constraint}) in the single brane model.

We can also express $\delta{\cal V}$ and $\delta {\cal C}$ in terms of $u_-(x)$ and $v_-(x)$.
With this we find
\begin{eqnarray}
\varphi \propto \delta {\cal V} \propto \delta{\cal C}.
\end{eqnarray}
This means that
the perturbation of the volume of the internal space and that of the circumference of the brane
are not independent, and $\varphi$ corresponds to this mode.

%
\begin{figure}[tb]
  \begin{center}
    \includegraphics[keepaspectratio=true,height=70mm]{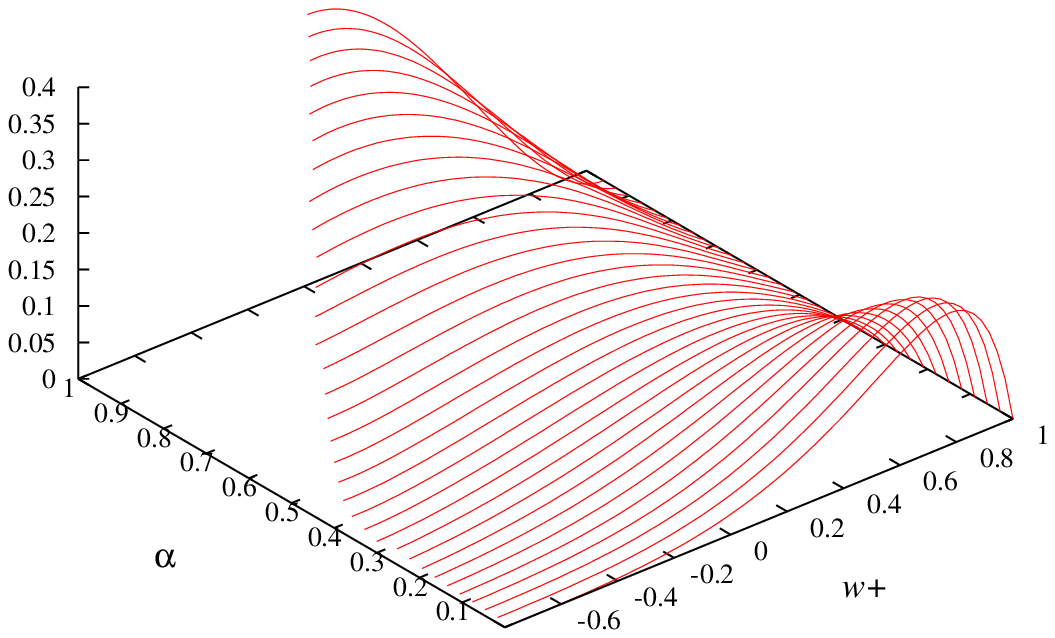}
  \end{center}
  \caption{$\Delta^{(+)}$ as a function of $\bar w_+$ and $\alpha$. The other parameters are fixed
  as $\bar w_-=-0.8$ and $\beta_0=0.8$. The domain of $\bar w_+$ is restricted
  from the condition $\bar w_+\geq$ max$\{\bar w_-, w_c\}$.}%
  \label{fig:two_D1_p.eps}
\end{figure}

\begin{figure}[tb]
  \begin{center}
    \includegraphics[keepaspectratio=true,height=70mm]{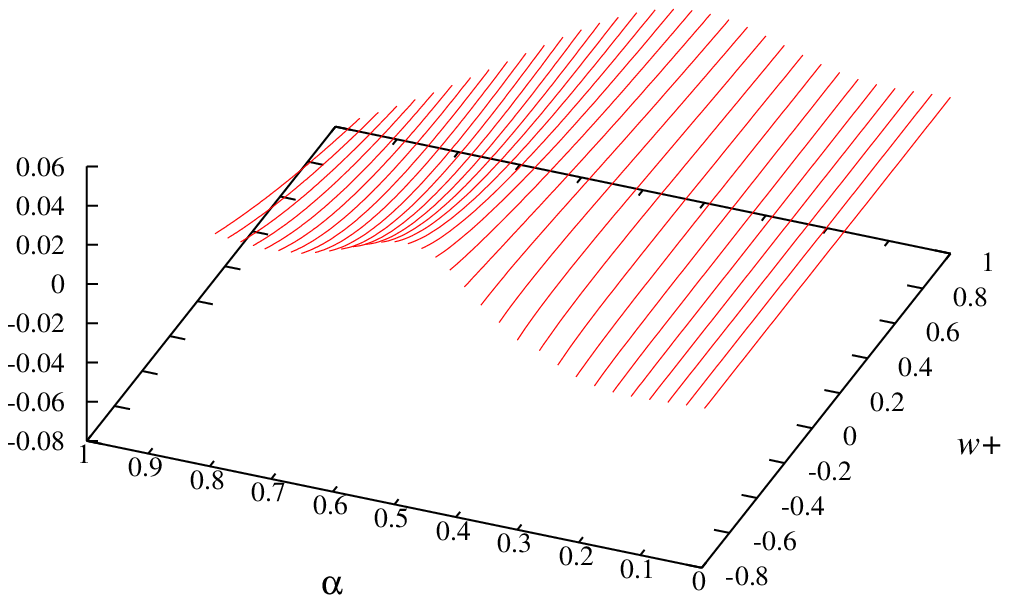}
  \end{center}
  \caption{$\Delta^{(-)}$ as a function of $\bar w_+$ and $\alpha$. The other parameters are fixed
  as $\bar w_-=-0.8$ and $\beta_0=0.8$. The domain of $\bar w_+$ is restricted
  from the condition $\bar w_+\geq$ max$\{\bar w_-, w_c\}$.}%
  \label{fig:two_D2_p.eps}
\end{figure}

\begin{figure}[tb]
  \begin{center}
    \includegraphics[keepaspectratio=true,height=70mm]{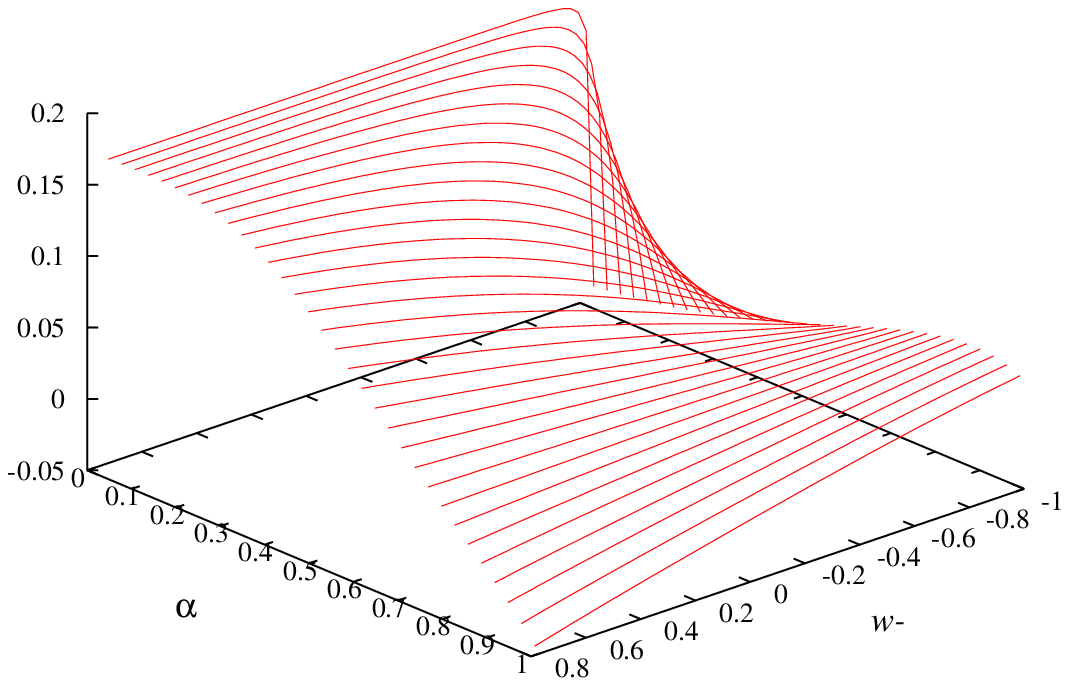}
  \end{center}
  \caption{$\Delta^{(+)}$ as a function of $\bar w_-$ and $\alpha$. The other parameters are fixed
  as $\bar w_+=0.8$ and $\beta_0=0.6$.}%
  \label{fig:two_D1_m.eps}
\end{figure}

\begin{figure}[tb]
  \begin{center}
    \includegraphics[keepaspectratio=true,height=70mm]{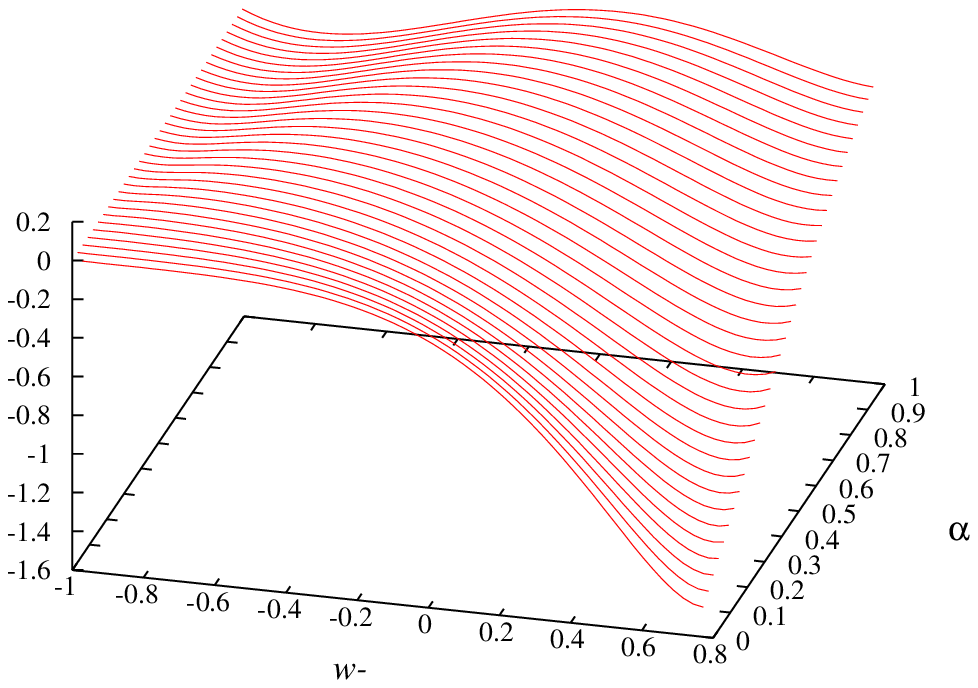}
  \end{center}
  \caption{$\Delta^{(-)}$ as a function of $\bar w_-$ and $\alpha$. The other parameters are fixed
  as $\bar w_+=0.8$ and $\beta_0=0.6$.}%
  \label{fig:two_D2_m.eps}
\end{figure}


The situation for the two-brane model is similar to that of the above single brane case.
There are $4\times3=12$ integration constants and
$2\times2=4$ brane bending scalars, while we have $4\times2=8$ boundary conditions
at the branes and $4$ regularity conditions at the poles.
Thus we can express $12$ of $16$ variables
in terms of $4$ variables, say $u_0(x)$, $v_0(x)$, $u_-(x)$, and $v_-(x)$.
Then Eqs.~(\ref{bdsc}) and~(\ref{evaluate_tau}) allow us to express
$\varphi^{(+)}$, $\bar\tau^{(+)}$, and $\bar\tau^{(-)}$ in terms of these four variables.
Interestingly, with this procedure
we can show that all the integration constants in $\varphi^{(+)}$ are encoded solely into $\bar\tau^{(\pm)}$ as
\begin{eqnarray}
\varphi^{(+)}=\frac{\ell_*^2}{a^2_+}8\pi G^{(+)}\left[\bar\tau^{(+)}
\Delta^{(+)}+\bar\tau^{(-)}
\Delta^{(-)}\right],
\label{p+dd}
\end{eqnarray}
where $\Delta^{(\pm)}$ are regular functions of
$\bar w_+, \bar w_-, \alpha$, and $\beta_0$.
Thus we arrive at the same conclusion as the above:
the effect of the mode $\varphi^{(+)}$ can be safely neglected on scales much larger than
$\ell_*$ on the brane.
Again, the explicit form of $\Delta^{(\pm)}$ is quite involved and so we just
illustrate their behavior in Figs.~\ref{fig:two_D1_p.eps}--\ref{fig:two_D2_m.eps}.
It turns out that $\Delta^{(+)}\to 0$
as the brane at $w=\bar w_+$ shrinks to the north pole, $\bar w_+\to +1$.
Similarly, we have $\Delta^{(-)}\to 0$ as the brane at $w=\bar w_-$ shrinks to the south pole, $\bar w_-\to-1$.
Thus, the $\varphi^{(+)}$ mode decouples from gravity on the brane when
one brings both branes to the poles.

It is straightforward to express
the perturbations of the circumferences of the two branes, $\delta{\cal C}^{(+)}$ and $\delta{\cal C}^{(-)}$,
and the volume fluctuation $\delta{\cal V}$, in terms of $u_0(x)$, $v_0(x)$, $u_-(x)$, and $v_-(x)$.
Comparing the result with $\bar\tau^{(\pm)}$ expressed in terms of these four variables,
we find that $\delta{\cal C}^{(\pm)}$
and $\delta{\cal V}$ can in fact be written as linear combinations of $\bar \tau^{(\pm)}$: 
\begin{eqnarray}
\delta{\cal C}^{(\pm)}&=&c^{(\pm)}_1\bar\tau^{(+)}+c^{(\pm)}_2\bar\tau^{(-)},
\\
\delta{\cal V}&=&d_1\bar\tau^{(+)}+d_2\bar\tau^{(-)}.
\end{eqnarray}
Namely, two of the three fluctuation modes $\delta{\cal C}^{(\pm)}$
and $\delta{\cal V}$ are linearly independent.
The mode $\varphi^{(+)}$ 
can be interpreted as a combination of two of these fluctuation modes.

Finally, let us take a brief look at the unwarped $\alpha=1$ model,
focusing on the simplest background configuration with $v_+=v_-$, i.e.,
the $Z_2$-symmetric configuration of the branes ($\bar w_-=-\bar w_+$).
In this case it is instructive to rearrange Eq.~(\ref{p+dd}) as
\begin{eqnarray}
\varphi^{(+)}=8\pi G^{(+)}\ell^2_*\left[\frac{\Delta^{(+)}+\Delta^{(-)}}{2}
\left(\bar\tau^{(+)}+\bar\tau^{(-)}\right)
+\frac{\Delta^{(+)}-\Delta^{(-)}}{2}
\left(\bar\tau^{(+)}-\bar\tau^{(-)}\right)\right].\label{nonz2p}
\end{eqnarray}
The coefficients are given by
\begin{eqnarray}
\Delta^{(+)}+\Delta^{(-)}&=&\frac{(1-\bar w_+)\left[9\beta+(10-7\beta)\bar w_+-(11+\beta)\bar w_+(1+\bar w_+)\right]}
{12\left[\beta(1-\bar w_+)+\bar w_+\right]},
\\
\Delta^{(+)}-\Delta^{(-)}&=&\frac{\bar w_+(1-\bar w_+)(10+\bar w_++\bar w_+^2)}
{2\left[ 10\beta+9(1-\beta)\bar w_++(1-\beta)\bar w_+^3 \right]}.
\end{eqnarray}
The restriction to $Z_2$-symmetric perturbations projects out the second term,
and then the result is in agreement with~\cite{Peloso1}.
In other words, Eq.~(\ref{nonz2p}) generalizes~\cite{Peloso1} to non-$Z_2$-symmetric perturbations.


\section{Conclusions}

In this paper we have studied linearized gravity on a brane in
a 6D Einstein-Maxwell system.
In the present braneworld model
two extra dimensions are compactified by a magnetic flux,
and hence the model is thought of as a toy model of string flux compactifications.
Introducing
codimension one branes (extended branes)
with one spatial dimension compactified on a Kaluza-Klein circle
instead of strict codimension two defects, one can put arbitrary energy-momentum tensor
on branes~\cite{Peloso1, PPZ, Peloso2}.
With this we can first discuss the behavior of gravity sourced by matter on the branes.
We have considered an axisymmetric, warped background with one or two branes,
and examined axisymmetric perturbations in this system,
generalizing the previous analysis~\cite{Peloso1} to the warped model of~\cite{PPZ}.
We have focused on the zero-mode sector of perturbations as the effect of
Kaluza-Klein modes is expected to be subdominant at long distances.

It is a well-known fact that the contribution from the brane bending mode is crucial
for recovering the 4D tensor structure
in the Randall-Sundrum braneworld~\cite{GT}.
We found that the same mechanism works in the present class of 6D warped braneworlds as well,
which has been shown only for the unwarped $Z_2$-symmetric
model in~\cite{Peloso1}.
There appears another scalar mode associated to the fluctuations
of the circumferences of the branes, so that
gravity is described by a scalar-tensor type theory.
However, we have shown that, as in the unwarped case, the scalar mode can be neglected
at distances much larger than the scale of flux compactification,
$\ell_* \sim \ell$.
Therefore, 4D Einstein gravity is reproduced on an extended brane in
6D warped flux compactifications.

On scales about $\ell$ the effect of Kaluza-Klein modes will not be negligible
in addition to the correction from the above scalar mode.
The precise evaluation of the corrections from the Kaluza-Klein modes
is important for testing the model against gravitational experiments and observations, which is one of our remaining issues.
Note, however, that the scale of the circumference of the brane is given by $\ell f^{1/2}(\bar w_i)$,
which is much smaller than the scale of the internal space $\ell$ provided that the location of the brane is
sufficiently close to a pole.
Consequently, the effect of the modes which are inhomogeneous
along the $\phi$ direction will become significant on much smaller scales than $\ell$.
Note also that even though we have one microscopic Kaluza-Klein direction,
the 4D Planck scale is given by Eq.~(\ref{4dpl}) [not by $M_{{\rm Pl}}\sim \ell\cdot\ell f^{1/2}(\bar w_i)M^4$].
Thus it is possible to have the scale $\ell$ of order submillimeter while addressing the hierarchy problem.

Finally we would like to make several remarks
on possible extensions of our current work.
First, the stability analysis for the present model remains to be done.
It would be also interesting to
explore cosmology and nonlinear brane gravity on an extend brane,
as we are now able to put arbitrary matter on the brane.
The cosmic expansion can be described by a moving brane
in a warped bulk, with the scale factor mimicked by the warp factor~\cite{branecos}.
However, in the present model the warp factor is bounded above
and hence it seems difficult to get an ever-expanding brane universe
following this line.
Note that a de Sitter universe on an extended brane has been constructed straightforwardly for
the unwarped bulk~\cite{Peloso2}. (See also Ref.~\cite{Mukohyama}.)
Going beyond the Einstein-Maxwell model, we also
plan to study aspects of gravity in the supersymmetric version
of the present brane model \cite{PPZ, Aghababaie:2003ar, 6d_sugra}.


\section*{Acknowledgements}

TK is supported by the JSPS under Contract No.~01642.
This work was also supported by the project ``Transregio (Dark
Universe)" at the ASC.



\appendix
\section{Derivation of the background metric}\label{App:deri}

We consider the 6D Einstein-Maxwell system described by the action
\begin{eqnarray}
S = \int d^6x\sqrt{-g}\left[\frac{M^4}{2}\left(\cR-\frac{1}{L^2}\right)-\frac{1}{4}\cF^2\right].
\label{app:ac}
\end{eqnarray}
The field equations derived from this action are
\begin{eqnarray}
\cR_{MN}-\frac{1}{2}g_{MN}\cR=-\frac{1}{2L^2}g_{MN}+\frac{1}{M^4}\left(
\cF_{ML}\cF_N^{~~L}-\frac{1}{4}g_{MN}\cF^2
\right).
\end{eqnarray}
We consider a solution obtained by a double Wick rotation from the
Reissner-Nordstr\"{o}m solution \cite{Mukohyama, Yoshiguchi, Sendouda}.
The metric can be written as
\begin{eqnarray}
g_{MN}dx^Mdx^N& =&\rho^2
\eta_{\mu\nu}d\tilde x^{\mu}d\tilde x^{\nu}+\frac{d\rho^2}{\tilde f(\rho)}+
\tilde f(\rho)d\tilde\phi^2,
\nonumber\\
\tilde f(\rho)&=&
\frac{D}{\rho^3}-\frac{\rho^2}{20L^2}-\frac{Q^2}{12M^4}\frac{1}{\rho^6},
\end{eqnarray}
and the field strength is given by
\begin{eqnarray}
\cF_{\rho\tilde\phi}=\frac{Q}{\rho^4}.
\end{eqnarray}

We assume that the metric function
$\tilde f(\rho)$ has two positive roots $\rho_+$ and $\rho_-$
and $\tilde f(\rho)>0$ for $\rho_-<\rho<\rho_+$.
The integration constants $D$ and $Q$ are conveniently parameterized by $\alpha:=\rho_-/\rho_+(\leq 1)$ as \cite{PPZ}
\begin{eqnarray}
D=\frac{\rho_+^5}{20L^2}\frac{1-\alpha^8}{1-\alpha^3},\quad
Q^2=\frac{\rho_+^8M^4}{L^2}\frac{3\alpha^3(1-\alpha^5)}{5(1-\alpha^3)}.
\end{eqnarray}
Introducing a new coordinate $w$ defined by
\begin{eqnarray}
a(w):= \frac{1}{2}\left[(1-\alpha)w+1+\alpha \right]
=\frac{\rho}{\rho_+},
\end{eqnarray}
we can rewrite the metric and the field strength as
\begin{eqnarray}
g_{MN}dx^Mdx^N&=&a^2\rho_+^2\eta_{\mu\nu}d\tilde x^{\mu}d\tilde x^{\nu}
+\frac{L^2dw^2}{f}+f\cdot\left[\frac{\rho_+}{2L}(1-\alpha)\right]^2d\tilde\phi^2,
\\
\cF_{\rho\tilde\phi}&=&\frac{1}{a^4}\frac{M^2}{L}
\sqrt{\frac{3\alpha^3(1-\alpha^5)}{5(1-\alpha^3)}},
\end{eqnarray}
where
\begin{eqnarray}
f=\frac{1}{5(1-\alpha)^2}\left[-a^2
+\frac{1-\alpha^8}{1-\alpha^3}\frac{1}{a^3}
-\frac{\alpha^3(1-\alpha^5)}{1-\alpha^3}\frac{1}{a^6}
\right].
\end{eqnarray}
Note here that $f$ depends only on the parameter $\alpha$.
The rescaling of the coordinates
\begin{eqnarray}
x^{\mu}=\rho_+\tilde x^{\mu},\quad
\beta\phi &=& \frac{\rho_+}{L^2}(1-\alpha)\tilde\phi,
\end{eqnarray}
leads to the background solution in the main text.

We would like to remark that in the above
the Minkowski metric $\eta_{\mu\nu}$ can be replaced
by any Ricci flat metric $g_{\mu\nu}^{(4)}$.


\section{Vector modes}\label{App:vcmode}

Under a vector gauge transformation
\begin{eqnarray}
x^{\mu}&\to& x^{\mu}+\hat\xi^{\mu},
\end{eqnarray}
the metric perturbations transform as
\begin{eqnarray}
E_{\mu}&\to&E_{\mu}-\hat\xi_{\mu},\nonumber\\
B_{w\mu}&\to&B_{w\mu}-a^2\hat\xi_{\mu}',\label{vectr}\\
B_{\mu}&\to&B_{\mu}.\nonumber
\end{eqnarray}
From Eqs.~(\ref{vectr}) we find the two gauge invariant variables
\begin{eqnarray}
V_{\mu}:=B_{w\mu}-a^2E_{\mu}',
\end{eqnarray}
and
$B_{\mu}$.
The perturbed gauge field $\hat A_{\mu}$ is also gauge invariant.

The linearized Einstein equations give
\begin{eqnarray}
V_{\mu}'+2\frac{a'}{a}V_{\mu}+\frac{f'}{f}V_{\mu}&=&0,
\label{v1}
\\
\Box V_{\mu}&=&0,
\label{v2}
\\
B_{\mu}''+2\frac{a'}{a}B_{\mu}'-6\left(\frac{a'}{a}\right)^2B_{\mu}
+\frac{L_I^2}{a^2f}\Box B_{\mu}&=&
-\frac{\ell}{M^2}\frac{2\mu}{a^4}\hat A_{\mu}'.
\label{v3}
\end{eqnarray}
From the $\mu$ component of the linearized Maxwell equations we obtain
\begin{eqnarray}
\left(a^2f\hat A_{\mu}'-\frac{M^2}{\ell}\frac{\mu}{a^2}B_{\mu}\right)'+L_I^2\Box B_{\mu}=0.
\label{v4}
\end{eqnarray}
The $(\mu\nu)$ component of the Israel conditions gives
\begin{eqnarray}
\frac{a^2_i f^{1/2}_i}{2\ell}
\left[\left[
\beta_I\left(V_{\mu,\nu}+V_{\nu,\mu}\right)
\right]\right]_{\bar w_i}
=
\frac{1}{M^4}T_{\mu\nu}^{(i)},
\label{v_Is_1}
\end{eqnarray}
where $a_i:=a(\bar w_i)$ and $f_i:=f(\bar w_i)$.
From the $(\mu\phi)$ component of the Israel conditions we have
\begin{eqnarray}
\left[\left[
\beta_I\left(\frac{1}{2}B_{\mu}'-\frac{a'}{a}B_{\mu}\right)\frac{f^{1/2}}{\ell}
\right]\right]_{\bar w_i}
=\left.\e v_i^2\left(\partial_{\phi}\sigma-\e\cA_{\phi}\right)\hat A_{\mu}\right|_{\bar w_i}.
\label{v_Is_2}
\end{eqnarray}
The Maxwell junction condition is
\begin{eqnarray}
\left[\left[\beta_I \hat A_{\mu}'\right]\right]_{\bar w_i}=\left.\ell (\e v_i)^2\hat A_{\mu}\right|_{\bar w_i}.
\label{v_Maxwell}
\end{eqnarray}

Equations~(\ref{v3}) and~(\ref{v4}) with~(\ref{v_Is_2}) and~(\ref{v_Maxwell})
govern the two vector variables $B_{\mu}$ and $\hat A_{\mu}$.
However, they do not couple to matter on the brane via the junction conditions.
For this reason we will not take care of these modes.

Equation~(\ref{v2}) indicates that only the zero mode is present for $V_{\mu}$.
Equation~(\ref{v1}) is then solved to give
\begin{eqnarray}
V_{\mu}(x, w)=\frac{{\mathtt V}^I_{\mu}(x)}{a^2f},
\end{eqnarray}
where ${\mathtt V}^I_{\mu}$ is an integration constant.
In order to ensure the regularity at the two poles, we impose
${\mathtt V}^I_{\mu}=0$ in the region that contains the pole. In the single brane model,
this immediately leads to $V_{\mu}=0$ everywhere.
In the absence of vector-type matter sources,
the source-free Israel condition, Eq.~(\ref{v_Is_1}) with $T_{\mu\nu}^{(i)}=0$,
forces $V_{\mu}$ to vanish everywhere in the two-brane model.


\section{Perturbation equations and gauges for scalar modes}\label{gts}

\subsection{Gauges}

Under a scalar gauge transformation
\begin{eqnarray}
x^{\mu}&\to& x^{\mu}+\xi^{,\mu},
\nonumber\\
w&\to& w+\xi^{w}, \label{gt}\\
\phi&\to& \phi+\xi^{\phi},\nonumber
\end{eqnarray}
the metric perturbations transform as
\begin{eqnarray}
\Psi&\to&\Psi-\frac{a'}{a} \xi^w,\nonumber\\
E&\to&E-\xi,\nonumber\\
B_w&\to&B_w-a^2\xi'-\frac{L_I^2}{f}\xi^w,\nonumber\\
B&\to&B-\ell^2f\xi^{\phi},\\
W&\to&W-{\xi^w}'+\frac{f'}{2f}\xi^w,\nonumber\\
\Phi&\to&\Phi-\frac{f'}{2f}\xi^w,\nonumber\\
C&\to&C-{\xi^{\phi}}'.\nonumber
\end{eqnarray}
The perturbed gauge field transforms as
\begin{eqnarray}
A&\to&A-\cA_\phi\xi^{\phi},\nonumber
\\
A_{w}&\to& A_{w}-\cA_{\phi}{\xi^{\phi}}',
\\
A_{\phi}&\to& A_{\phi}-\xi^w\cA_{\phi}',\nonumber
\end{eqnarray}
and so
\begin{eqnarray}
F_w\to F_w+\cA_{\phi}'\xi^{\phi}.
\end{eqnarray}

One of the gauge we employ in the present paper is an analogue to the longitudinal gauge,
which is defined by $E=B_w=B=0$. Hence we can define metric perturbations
in the longitudinal gauge as
\begin{eqnarray}
\tilde\Psi &:=& \Psi- \frac{a'}{a}\frac{f}{L_I^2}X,
\\
\tilde W&:=&W-\frac{f}{L^2}X'-\frac{f'}{2L^2_I}X,
\\
\tilde \Phi&:=&\Phi-\frac{f'}{2L^2_I}X,
\\
\tilde C&:=&C-\left(\frac{B}{\ell^2f}\right)',
\end{eqnarray}
where we defined a convenient variable $X:=B_w-a^2E'$.
Similarly, the gauge field perturbations in the longitudinal gauge can be defined as
\begin{eqnarray}
\tilde A&:=& A-\cA_{\phi}\left(\frac{B}{\ell^2f}\right),
\\
\tilde A_w&:=& A_w-\cA_{\phi}\left(\frac{B}{\ell^2f}\right)',
\\
\tilde A_{\phi}&:=& A_{\phi}-\cA_{\phi}'\frac{f}{L_I^2}X.
\end{eqnarray}
Scalar-type variables with tilde will denote perturbations in the longitudinal gauge.

The 6D perturbation equations
in the longitudinal gauge reduce to a system of two (coupled) master equations.
For this reason the longitudinal gauge is useful for solving the bulk equations. 
In this gauge, however, the positions of the branes $w_b^{(i)}$ are perturbed in general:
\begin{eqnarray}
w_b^{(i)}=\bar w_{i}+\zeta^{(i)}(x).
\end{eqnarray}
Strictly speaking,
we have two different brane bending modes for each
one brane, $\zeta^{(i)}_{w>\bar w_i}$ and $\zeta^{(i)}_{w<\bar w_i}$,
as no $Z_2$-symmetry is assumed at the branes.

The Gaussian-normal gauge,
defined by
\begin{eqnarray}
B_w=W=C=0 \quad\text{and}\quad w_{b}^{(i)}=\bar w_{i},
\end{eqnarray}
is convenient
when imposing the boundary conditions at the brane as
the brane position is not perturbed.
We can employ this gauge in the neighborhood of each brane.
Hereafter, scalar-type variables with bar will denote perturbations in the Gaussian-normal gauge.
Starting from the longitudinal gauge, the Gaussian-normal gauge is realized
by a coordinate transformation $\bar x^M=\tilde x^M+\xi^M$ such that
\begin{eqnarray}
0&=&-a^2\xi'-\frac{L^2_I}{f}\xi^w,
\\
0&=&\tilde W-{\xi^w}'+\frac{f'}{2f}\xi^w,
\\
0&=&\tilde C-{\xi^{\phi}}',
\\
0&=&\zeta^{(i)}+\xi^w|_{\bar w_{i}}.
\end{eqnarray}
We can fix the residual gauge freedom by imposing
$\xi|_{\bar w_{i}}=\xi^{\phi}|_{\bar w_{i}}=0$.
With this, perturbative matter sources localized on the brane
are invariant when going from the longitudinal gauge to the Gaussian-normal gauge.

The above gauge transformation relates the longitudinal gauge perturbations
with the perturbations of the induced metric and gauge field perturbations on the brane.
For example, we find
\begin{eqnarray}
\bar\Psi|_{\bar w_i}&=&\left.\left[\tilde\Psi+\frac{a'}{a}\zeta^{(i)}\right]\right|_{\bar w_i},
\\
\bar\Phi|_{\bar w_i}&=&\left.\left[\tilde\Phi+\frac{f'}{2f}\zeta^{(i)}\right]\right|_{\bar w_i}.
\end{eqnarray}
We also have $\bar A|_{\bar w_i}=\tilde A|_{\bar w_i}$ and
$\bar F_w|_{\bar w_i}=\tilde F_w|_{\bar w_i}$.

\subsection{6D perturbation equations for scalar modes}

We work in the longitudinal gauge for the analysis of the bulk perturbations.
The linearized Einstein equations give
\begin{eqnarray}
&&\frac{L_I^2}{a^2}\Box\left(\tilde\Phi+3\tilde\Psi\right)
+2f\left[2\frac{a'}{a} \tilde\Phi'+\left(\frac{f'}{f}+6 \frac{a'}{a}\right)\tilde\Psi'\right]
+\tilde W+\frac{\mu^2}{a^8}\tilde\Phi
=\frac{\mu}{\ell M^2a^4}\tilde A_{\phi}',
\label{ww}
\\
&&4f\left[\tilde\Psi''+\left(\frac{f'}{2f}+5\frac{a'}{a}\right)\tilde\Psi'
- \frac{a'}{a} \tilde W'+ \frac{a'}{a}\left(\frac{f'}{f}+3 \frac{a'}{a}\right)(\tilde\Phi-\tilde W)\right]
\nonumber\\&&\qquad\qquad\qquad\qquad\qquad\qquad 
+\frac{L_I^2}{a^2}\Box\left(3\tilde\Psi+\tilde W\right)+\tilde\Phi+\frac{\mu^2}{a^8}\tilde W=
\frac{\mu}{\ell M^2a^4}\tilde A_{\phi}',
\label{phiphi}
\\
&&f\left[\tilde\Phi'+3\tilde\Psi'+\left(\frac{f'}{2f}-\frac{a'}{a}\right)\tilde\Phi
-\left(\frac{f'}{2f}+3 \frac{a'}{a}\right)\tilde W
\right]
=-\frac{\mu}{\ell M^2a^4}\tilde A_{\phi},
\label{wmu}
\\
&&f\Biggl[\tilde\Phi''+3\tilde\Psi''-\left(\frac{f'}{2f}+3\frac{a'}{a}\right)\tilde W'
+3\left(\frac{f'}{2f}+\frac{a'}{a}\right)\tilde \Phi'+3\left(\frac{f'}{f}+4\frac{a'}{a}\right)\tilde\Psi'
\nonumber\\&&\qquad
+\left(\frac{f''}{f}+6\frac{a'}{a}\frac{f'}{f}+6\frac{a'^2}{a^2}\right)(\tilde\Psi-\tilde W)\Biggr]
-\frac{\mu^2}{a^8}(\tilde W+\tilde \Phi)+\left(1+\frac{\mu^2}{a^8}\right)\tilde \Psi
=-\frac{\mu}{\ell M^2a^4}\tilde A_{\phi}',
\label{tracelessmunu}
\\&&\qquad\qquad\qquad\qquad\qquad\qquad\qquad
2\tilde\Psi+\tilde\Phi+\tilde W=0, \label{tracemunu}
\end{eqnarray}
and
\begin{eqnarray}
\Box\tilde C&=&0,
\label{wphi}
\\
\tilde C'+2\left(\frac{f'}{f}+\frac{a'}{a} \right)\tilde C&=&-\frac{2\mu}{\ell M^2a^4}\tilde F_w.
\label{phimu}
\end{eqnarray}
From the linearized Maxwell equations $\partial_M\delta\left[\sqrt{-g}\cF^{MN}\right]=0$ we get
\begin{eqnarray}
\left[a^4\tilde A_{\phi}'+\mu M^2\ell(4\tilde\Psi-\tilde\Phi-\tilde W)\right]'
+\frac{a^2L_I^2}{f}\Box \tilde A_{\phi}=0,
\label{maxwell_phi}
\end{eqnarray}
and
\begin{eqnarray}
\left(a^2f\tilde F_w\right)'&=&0,
\label{maxwell_F1}\\
\Box \tilde F_w&=&0.
\label{maxwell_F2}
\end{eqnarray}

On each brane we have the perturbed equation of motion for $\sigma_i$ field:
$
\partial_{\hat\mu}\left[\sqrt{-q}q^{\hat\mu\hat\nu}\left(
\partial_{\hat\nu}\sigma_i-\e \cA_{\hat\nu}\right)\right]=0,
$
or, explicitly,
\begin{eqnarray}
\partial^{\phi}\partial_{\phi}\delta\sigma_i+\frac{1}{a^2}\Box\left(\delta\sigma_i
-\e \tilde A\right) =0.
\label{eom-sigma}
\end{eqnarray}

Note here that under an infinitesimal $U(1)$ gauge transformation,
$\delta\cA_M\to\delta\cA_M+\partial_M\Xi$ and $\delta\sigma_i\to\delta\sigma_i+\e\Xi$,
where $\Xi$ is independent of $\phi$,
the above equations of motion (and boundary conditions
in the main text) are invariant.


\section{On $\tilde C$ and $\tilde F_w$}\label{App:vm}

The continuity of the gauge field imposes
\begin{eqnarray}
\left[\left[\tilde A\right]\right]_{\bar w_i}&=&0.
\label{con4}
\end{eqnarray}
From the $(\phi\mu)$ component of the Israel conditions we obtain
\begin{eqnarray}
\left[\left[\beta_I\tilde C\right]\right]_{\bar w_i}
+[[\beta_I]]_{\bar w_i}\:\frac{2\mu}{\e\ell M^2a_i^4f_i}\left.\left(\delta\sigma_i-\e\tilde A\right)\right|_{\bar w_i}
=0,\label{pmuIS}
\end{eqnarray}
where $a_i:=a(\bar w_i)$ and $f_i:=f(\bar w_i)$.
From the $\mu$ component of the Maxwell junction conditions we get
\begin{eqnarray}
\left[\left[ \beta_I\tilde F_w\right]\right]_{\bar w_i}=
\frac{\ell\e v^2_i}{f_i^{1/2}}\left.\left(\delta\sigma_i-\e \tilde A\right)\right|_{\bar w_i}.
\label{mmu}
\end{eqnarray}
We can eliminate $(\delta\sigma_i-\e \tilde A)|_{\bar w_i}$ from Eqs.~(\ref{pmuIS}) and~(\ref{mmu}) to get
\begin{eqnarray}
a^4_if^{1/2}_i \left[\left[\beta_I\tilde C\right]\right]_{\bar w_i}
+\frac{2\mu}{(\e v_i\ell M)^2}
[[\beta_I]]_{\bar w_i}\left[\left[ \beta_I\tilde F_w\right]\right]_{\bar w_i}=0.
\label{juncCF}
\end{eqnarray}

Equations~(\ref{wphi}),~(\ref{phimu}),~(\ref{maxwell_F1}),~(\ref{maxwell_F2}),
and~(\ref{juncCF}) supplemented with appropriate boundary conditions
at the poles form a closed set of equations that can determine $\tilde C$ and $\tilde F_w$.
Then Eqs.~(\ref{eom-sigma}),~(\ref{con4}), and~(\ref{pmuIS}) [or (\ref{mmu})]
are enough to determine $\delta\sigma_i$ and $\tilde A$ on the brane.
In Appendix \ref{App:vm}, we show that $\tilde C$ and $\tilde F_w$ modes vanish everywhere
both for the single brane and two-brane models.

Equation~(\ref{wphi}) shows that only the zero mode is present
for $\tilde C$. The same is true for $\tilde F_w$, as is clear from Eq.~(\ref{maxwell_F2}).
The general solution to Eqs.~(\ref{phimu}) and~(\ref{maxwell_F1}) is given by
\begin{eqnarray}
\tilde C&=&\frac{{\mathtt C}^I(x)}{a^2f^2}-{\mathtt F}^I(x)\int^w\frac{f}{a^4}dw,
\\
\tilde F_w&=&\frac{\ell M^2}{2\mu} \frac{{\mathtt F}^I(x)}{a^2f},
\end{eqnarray}
where ${\mathtt C}^I$ and ${\mathtt F}^I$ are integration constants.
The regularity at the poles requires that ${\mathtt C}^{\pm}$ and ${\mathtt F}^{\pm}$ vanish.
Thus in the single brane model $\tilde C=0$ and $\tilde F_w=0$ everywhere.
In the two-brane model, we can write
\begin{eqnarray}
\tilde C&=&\frac{{\mathtt C}^0(x)}{a^2f^2}+{\mathtt F}^0(x)\int^{\bar w_+}_w\frac{f}{a^4}dw,
\\
\tilde F_w&=&\frac{\ell M^2}{2\mu} \frac{{\mathtt F}^0(x)}{a^2f},
\end{eqnarray}
for $\bar w_-<w<\bar w_+$.
However, the junction condition~(\ref{juncCF}) at both branes reads
\begin{eqnarray}
{\mathtt C}^0+\frac{\beta_+-\beta_0}{\ell (\e v_+)^2}
\left.\frac{f^{1/2}}{a^4}\right|_{\bar w_+}{\mathtt F}^0&=&0,
\\
{\mathtt C}^0+\left.\left[\left(\int^{\bar w_+}_{\bar w_-}\frac{f}{a^4}dw\right)a^2f^2
+\frac{\beta_0-\beta_-}{\ell(\e v_-)^2}\frac{f^{1/2}}{a^4}\right]\right|_{\bar w_-}{\mathtt F}^0&=&0,
\end{eqnarray}
leading to ${\mathtt C}^0=0$ and ${\mathtt F}^0=0$.
Thus we see that $\tilde C=0$ and $\tilde F_w=0$ everywhere also in the two-brane model.
Then from Eq.~(\ref{pmuIS}) or Eq.~(\ref{mmu}) we have
$\delta\sigma_i - \e \tilde A =0$,
at the branes. Eq.~(\ref{eom-sigma}) now reduces to $\partial^\phi\partial_\phi\delta \sigma_i=0$, 
which is solved by
$\delta\sigma_i=\delta\sigma_i^{(1)}(x)+\delta\sigma_i^{(2)}(x)\phi$.
However, $\delta\sigma^{(2)}_i=0$ because of periodicity in $\phi$ of $\sigma_i$ and the fact that $\delta\sigma_i$
is infinitesimal. Thus we get
$\delta\sigma_i = \e \tilde A|_{\bar w_i} = \delta \sigma_i^{(1)}(x)$.





\begin{thebibliography}{99}


\bibitem{Rubakov}
  See, e.g., V.~A.~Rubakov,
  Phys.\ Usp.\  {\bf 44}, 871 (2001)
  [Usp.\ Fiz.\ Nauk {\bf 171}, 913 (2001)]
  [arXiv:hep-ph/0104152].

\bibitem{Arkani-Hamed:1998rs}
  N.~Arkani-Hamed, S.~Dimopoulos and G.~R.~Dvali,
  Phys.\ Lett.\  B {\bf 429}, 263 (1998)
  [arXiv:hep-ph/9803315].


\bibitem{Papantonopoulos:2006uj}
  E.~Papantonopoulos,
  arXiv:gr-qc/0601011.


\bibitem{cc}
J.~W.~Chen, M.~A.~Luty and E.~Ponton,
  JHEP {\bf 0009}, 012 (2000)
  [arXiv:hep-th/0003067];
   S.~M.~Carroll and M.~M.~Guica,
  arXiv:hep-th/0302067;
  I.~Navarro,
  JCAP {\bf 0309}, 004 (2003)
  [arXiv:hep-th/0302129];
 Y.~Aghababaie, C.~P.~Burgess, S.~L.~Parameswaran and F.~Quevedo,
  Nucl.\ Phys.\  B {\bf 680}, 389 (2004)
  [arXiv:hep-th/0304256].



\bibitem{cc2}
I.~Navarro,
  Class.\ Quant.\ Grav.\  {\bf 20}, 3603 (2003)
  [arXiv:hep-th/0305014];
  H.~P.~Nilles, A.~Papazoglou and G.~Tasinato,
  Nucl.\ Phys.\  B {\bf 677}, 405 (2004)
  [arXiv:hep-th/0309042];
  H.~M.~Lee,
  Phys.\ Lett.\  B {\bf 587}, 117 (2004)
  [arXiv:hep-th/0309050];
  J.~Vinet and J.~M.~Cline,
  Phys.\ Rev.\  D {\bf 70}, 083514 (2004)
  [arXiv:hep-th/0406141];
J.~Garriga and M.~Porrati,
  JHEP {\bf 0408}, 028 (2004)
  [arXiv:hep-th/0406158].




\bibitem{Mukohyama}
  S.~Mukohyama, Y.~Sendouda, H.~Yoshiguchi and S.~Kinoshita,
  JCAP {\bf 0507}, 013 (2005)
  [arXiv:hep-th/0506050].







\bibitem{pt}
J.~M.~Cline, J.~Descheneau, M.~Giovannini and J.~Vinet,
  JHEP {\bf 0306}, 048 (2003)
  [arXiv:hep-th/0304147];
  O.~Corradini, A.~Iglesias, Z.~Kakushadze and P.~Langfelder,
  Mod.\ Phys.\ Lett.\  A {\bf 17}, 795 (2002)
  [arXiv:hep-th/0201201].





\bibitem{Bostock}
  P.~Bostock, R.~Gregory, I.~Navarro and J.~Santiago,
  Phys.\ Rev.\ Lett.\  {\bf 92}, 221601 (2004)
  [arXiv:hep-th/0311074];


  \bibitem{shock}
  N.~Kaloper and D.~Kiley,
  JHEP {\bf 0603}, 077 (2006)
  [arXiv:hep-th/0601110].


\bibitem{Peloso1}
  M.~Peloso, L.~Sorbo and G.~Tasinato,
  Phys.\ Rev.\ D {\bf 73}, 104025 (2006)
  [arXiv:hep-th/0603026].



\bibitem{Corradini}
  O.~Corradini and Z.~Kakushadze,
  Phys.\ Lett.\  B {\bf 506}, 167 (2001)
  [arXiv:hep-th/0103031];
  O.~Corradini, A.~Iglesias, Z.~Kakushadze and P.~Langfelder,
  Phys.\ Lett.\  B {\bf 521}, 96 (2001)
  [arXiv:hep-th/0108055].

\bibitem{ish}
  I.~P.~Neupane,
  Class.\ Quant.\ Grav.\  {\bf 19}, 5507 (2002)
  [arXiv:hep-th/0106100];
  Y.~M.~Cho and I.~P.~Neupane,
  Int.\ J.\ Mod.\ Phys.\  A {\bf 18}, 2703 (2003)
  [arXiv:hep-th/0112227].


\bibitem{Kaloper:2004cy}
  N.~Kaloper,
  JHEP {\bf 0405}, 061 (2004)
  [arXiv:hep-th/0403208].



\bibitem{Kanno:2004nr}
  S.~Kanno and J.~Soda,
  JCAP {\bf 0407}, 002 (2004)
  [arXiv:hep-th/0404207].

\bibitem{Charmousis:2005ey}
  C.~Charmousis and R.~Zegers,
  JHEP {\bf 0508}, 075 (2005)
  [arXiv:hep-th/0502170];
  Phys.\ Rev.\  D {\bf 72}, 064005 (2005)
  [arXiv:hep-th/0502171].




\bibitem{deRham:2005ci}
  C.~de Rham and A.~J.~Tolley,
  JCAP {\bf 0602}, 003 (2006)
  [arXiv:hep-th/0511138].







\bibitem{Carter:2006uk}
J.~Louko and D.~L.~Wiltshire,
  JHEP {\bf 0202}, 007 (2002)
  [arXiv:hep-th/0109099];
  B.~M.~N.~Carter, A.~B.~Nielsen and D.~L.~Wiltshire,
  JHEP {\bf 0607}, 034 (2006)
  [arXiv:hep-th/0602086].

\bibitem{Kaloper}
  N.~Kaloper,
  arXiv:hep-th/0702206.

\bibitem{Peloso2}
  B.~Himmetoglu and M.~Peloso,
  arXiv:hep-th/0612140.

\bibitem{PPZ}
  E.~Papantonopoulos, A.~Papazoglou and V.~Zamarias,
  arXiv:hep-th/0611311.









\bibitem{Aghababaie:2003ar}
  Y.~Aghababaie {\it et al.},
  JHEP {\bf 0309}, 037 (2003)
  [arXiv:hep-th/0308064].



\bibitem{6d_sugra}
  G.~W.~Gibbons, R.~Guven and C.~N.~Pope,
  Phys.\ Lett.\  B {\bf 595}, 498 (2004)
  [arXiv:hep-th/0307238];
  C.~P.~Burgess, F.~Quevedo, G.~Tasinato and I.~Zavala,
  JHEP {\bf 0411}, 069 (2004)
  [arXiv:hep-th/0408109].







\bibitem{GT}
  J.~Garriga and T.~Tanaka,
  Phys.\ Rev.\ Lett.\  {\bf 84}, 2778 (2000)
  [arXiv:hep-th/9911055].




\bibitem{Randall:1999ee}
  L.~Randall and R.~Sundrum,
  Phys.\ Rev.\ Lett.\  {\bf 83}, 3370 (1999)
  [arXiv:hep-ph/9905221];
  Phys.\ Rev.\ Lett.\  {\bf 83}, 4690 (1999)
  [arXiv:hep-th/9906064].



  
\bibitem{FR}
  P.~G.~O.~Freund and M.~A.~Rubin,
  Phys.\ Lett.\  B {\bf 97}, 233 (1980);
  S.~Randjbar-Daemi, A.~Salam and J.~A.~Strathdee,
  Nucl.\ Phys.\  B {\bf 214}, 491 (1983).






\bibitem{Yoshiguchi}
  H.~Yoshiguchi, S.~Mukohyama, Y.~Sendouda and S.~Kinoshita,
  JCAP {\bf 0603}, 018 (2006)
  [arXiv:hep-th/0512212].

\bibitem{Sendouda}
  Y.~Sendouda, S.~Kinoshita and S.~Mukohyama,
  Class.\ Quant.\ Grav.\  {\bf 23}, 7199 (2006)
  [arXiv:hep-th/0607189].

\bibitem{branecos}
H.~A.~Chamblin and H.~S.~Reall,
  Nucl.\ Phys.\  B {\bf 562}, 133 (1999)
  [arXiv:hep-th/9903225];
P.~Kraus,
  JHEP {\bf 9912}, 011 (1999)
  [arXiv:hep-th/9910149];
A.~Kehagias and E.~Kiritsis,
  JHEP {\bf 9911}, 022 (1999)
  [arXiv:hep-th/9910174];
  D.~Ida,
  JHEP {\bf 0009}, 014 (2000)
  [arXiv:gr-qc/9912002].
  
  

  


\end{thebibliography}
\end{document}